\documentclass[a4paper]{article}
\usepackage{graphicx}
\usepackage{twocolceurws}
\usepackage[numbers]{natbib}
\usepackage{titlesec}
\usepackage{enumitem}
\usepackage{booktabs}
\usepackage{float}
\usepackage{cleveref}
\usepackage{multirow}
\usepackage{todonotes}
\usepackage{tikz}
\usepackage{caption}
\usepackage{natbib}

\title{The Impact of AI Explanations on Clinicians’ Trust and Diagnostic Accuracy in Breast Cancer}

\author{
Olya Rezaeian \\ Dept. of Systems and Enterprises\\
                Stevens Institute of Technology   
\and
Onur Asan \\ Dept. of Systems and Enterprises\\
                Stevens Institute of Technology    
\and
\\
Alparslan Emrah Bayrak \\ Dept. of Mechanical Engineering and Mechanics\\ Lehigh University
}

\institution{}
\makeatletter
\def\copyrightspace{} 
\gdef\@thanks{} 
\makeatother
\begin{document}

\maketitle

\begin{abstract}
Advances in machine learning have created new opportunities to develop artificial intelligence (AI)-based clinical decision support systems using past clinical data and improve diagnosis decisions in life-threatening illnesses such breast cancer. Providing explanations for AI recommendations is a possible way to address trust and usability issues in black-box AI systems. This paper presents the results of an experiment to assess the impact of varying levels of AI explanations on clinicians' trust and diagnosis accuracy in a breast cancer application and the impact of demographics on the findings. The study includes 28 clinicians with varying medical roles related to breast cancer diagnosis. The results show that increasing levels of explanations do not always improve trust or diagnosis performance. The results also show that while some of the self-reported measures such as AI familiarity depend on gender, age and experience, the behavioral assessments of trust and performance are independent of those variables.
\end{abstract}

\section{Introduction}

Given the data-intensive nature of healthcare, AI-based Clinical Decision Support Systems (CDSS) have become increasingly prevalent across various domains, including diagnosis \cite{mckinney_international_2020,micocci_attitudes_2021, janowczyk_deep_2016, nahata_deep_2020}, treatment planning\cite{mcintosh_voxel-based_2016, choudhury_effect_2022}, and symptom checking \cite{woodcock_impact_2021}. CDSS systems offer numerous advantages, such as reducing diagnostic errors, improving efficiency, and lowering costs if designed and utilized efficiently \cite{cartolovni_ethical_2022,gaube_as_2021,topol_high-performance_2019}. Many AI systems have been developed to enhance the prediction of clinical outcomes, especially for breast cancer detection \cite{sholehrasa_integrating_2024}. However, despite their potential, one of the most significant challenges remains the establishment of factors that influence the interaction between clinicians and these AI-driven tools, such as trust, perceived usefulness, and ease of use. No matter how advanced or effective a CDSS may be, its success attaches to clinicians' willingness to use and integrate its recommendations into their practice \cite{parasuraman_humans_1997}. Thus, trust is a fundamental factor in successfully adopting and utilizing AI in healthcare settings \cite{choudhury_effect_2022}. 

Trust in AI systems is influenced by numerous factors, as highlighted by a substantial body of research. Among these, the explainability of AI systems has emerged as one of the most critical elements \cite{tucci_factors_2022}, particularly due to the "black-box" nature of many machine learning models, which can make it difficult for clinicians to understand how decisions are reached. It is known in the literature that the type and medium of information presented to the clinicians impact their tendency to agree with the system \cite{miller_effects_2009}. However, the exact role that explainability plays in shaping clinician behavior is not yet fully understood.

In this experimental study, we explore the interaction between clinicians and our AI-based CDSS for breast cancer diagnosis, focusing on the impact of varying levels of explainability. Our primary aim is to evaluate how different degrees of AI explainability influence clinicians' diagnostic accuracy, as well as their self-reported and behavioral trust in the system. To achieve this, we present the results of a human-subject experiment using a custom-built AI web application that provides diagnostic recommendations generated through machine learning techniques with varying levels of explainability in multiple interventions. 

The experimental data we present includes the clinicians' interactions with system and their diagnostic decisions as well as pre- and post-experiment surveys assessing demographics and participants' perceptions and trust levels. Data from these experiments enable us to compare outcomes across different levels of explainability in multiple interventions to determine the optimal balance that fosters trust, adoption, and effective decision-making. The findings provide valuable insights into best practices for integrating explainability into AI systems in healthcare, ultimately contributing to more informed, data-driven clinical decisions and improved patient outcomes.

Aligned with the objectives of this study, we seek to address the following research questions:
\begin{itemize}
    \item{\bfseries\textit{RQ1: }}Does providing explanations improve both human decision-making performance and trust in AI-based clinical decision support systems for breast cancer detection?
    \item{\bfseries\textit{RQ2: }}How do clinicians' demographic characteristics, such as age, gender, and familiarity with AI, influence their trust in and performance with AI-based clinical decision support systems (CDSS) in breast cancer detection?
\end{itemize}

\section{Related Work}

\subsection{Trust in Human-Technology Interaction}
Trust plays a key role in how much users rely on a system, influencing their willingness to use it \cite{lee_trust_2004, niu_relationship_2018, choudhury_effect_2022, kim_factors_2020}. There are many studies that extensively explore how establishing trust encourages users to depend on a system, ultimately shaping their decisions to engage with it. Muir's work \cite{muir_trust_1987} serves as a valuable foundation for understanding human-machine interactions, proposing a trust model that explores various aspects of this relationship and provides a basis for testing future hypotheses. One of the earliest and most influential models of trust was introduced by Mayer et al., who proposed ability, benevolence, and integrity as factors affecting trustworthiness \cite{mayer_integrative_1995}. However, Mayer's model primarily focused on interpersonal trust. In the context of human-technology interaction, Parasuraman et al. \cite{parasuraman_humans_1997} expanded this understanding by examining how users interact with automation technology. Their study explored aspects such as usage patterns, reliance, and potential for misuse or overreliance on automation. They identified factors like trust, workload, and perceived risk as significant in determining whether individuals choose to engage with automated systems. Parasuraman et al. also introduced the concept of automation abuse, where system designs overly prioritize automation, potentially leading to diminished human oversight and performance. Such an approach can lead to both overreliance and neglect of automation by human operators. This finding is supported by other research, such as \cite{asan_artificial_2020}, which emphasizes that overreliance on automation may result in errors and negative consequences, including insufficient system monitoring or biased decision-making. Thus, the studies on literature highlights the need for designing the technology for appropriate reliance or "calibrated trust" ~\cite{asan_artificial_2020, lee_trust_2004}. Chiou \cite{chiou_trusting_2023} studies trust in workplace automation by exploring the interplay between humans and automation beyond traditional trust calibration. The study emphasizes responsive automation to enhance collaboration, aiming not just for trust calibration but for cultivating trust through the adaptability of automation systems.

To build upon these insights, it is essential to recognize that measuring trust in technology is a complex task. Trust is often conceptualized as a latent variable, meaning it cannot be directly observed but must instead be inferred through indirect measures \cite{kohn_measurement_2021}. To address this, Kohn et.al \cite{kohn_measurement_2021} proposed three main categories of measures that can be used to assess trust: self-reported measures, behavioral measures, and physiological measures. In this study, we will focus on utilizing the first two — self-reported and behavioral measures — to assess clinicians' trust in AI-based systems.

\subsection{Factors Affecting Trust Formation in Technology}
Many researchers have explored factors influencing users' trust, aiming to enhance trust in AI systems for effective human-AI collaboration. Hoff and Bashir \cite{hoff_trust_2015} propose that the trust formation process comprises three layers: dispositional trust, situational trust, and learned trust. Building on those layers, researchers have identified user-related, AI-related, and wider contextual factors that influence trust formation between users and technological systems~\cite{jermutus_influences_2022, lotfalian_saremi_survey_2021}.

For example, McIntosh et al. \cite{mcintosh_clinical_2021} observed a decline in real-world adoption of AI-based radiation therapy plans despite demonstrated efficiency, highlighting a gap between trust in simulation and clinical use. Similarly, in spacecraft rendezvous and docking (RVD), automation improved performance for both novices and experts, but trust varied, with experts' trust linked to specific performance metrics such as control error \cite{niu_relationship_2018}. Additionally, a study on the Blood Utilization Calculator (BUC) \cite{choudhury_effect_2022} emphasized that trust is influenced by perceived risk and expectancy, shaping the intent to use AI in healthcare. In human-robot interaction(HRI), a structural equation model (SEM) indicated the interrelationship of factors affect trust formation \cite{kim_factors_2020}.

Among the several factors influencing users' trust in AI, explainability has consistently been highlighted as one of the most important elements \cite{tucci_factors_2022, shin_effects_2021, bernardo_affective_2023}. The significance of explainability emphasizes the need for clear and transparent communication of AI decision-making processes to users. When users understand how AI systems arrive at their conclusions, it helps build trust and confidence in the technology \cite{jermutus_influences_2022}.

\subsection{The Effect of Explainability on Trust}

With the development of machine learning models, their complexity has increased, resulting in what is referred to as "black-box" models, where understanding how a model reaches a decision is challenging \cite{carvalho_machine_2019}. To address this, a range of post-hoc techniques have been developed to provide explanations for these opaque models. Researchers are increasingly exploring the effectiveness of these explainable AI (commonly referred to as xAI) methods in decision-making processes. 

XAI methods can be broadly categorized into global and local explanations, depending on whether they explain the model's behavior for an individual prediction or across the entire dataset. Global explanation methods provide insights into how the model behaves as a whole, helping users understand the general patterns and rules that the model follows. Examples of such methods include SHAP when aggregated over multiple instances to reveal global trends \cite{lundberg_unified_2017}, Concept Attribution, and Trust Score, which offer a holistic view of model reliability.

On the other hand, local explanation methods focus on explaining a model’s decision for a specific instance or a small subset of data. LIME \cite{ribeiro__2016} and SHAP in its local application provide such instance-specific explanations. Other local methods include Attribution-based methods like Saliency Maps  and Counterfactual explanations, which help users understand how individual features influenced a specific prediction. Prototype and Example-based techniques also contribute to local interpretability by comparing predictions to similar cases or providing representative examples.

Recent studies have explored the role of these methods in enhancing trust and performance in AI-assisted tasks, with varying results. Wang et al. \cite{wang_rationality_2024} investigated the use of SHAP in sales prediction and found that while SHAP-enhanced explanations improved decision accuracy and behavioral trust, they did not significantly affect self-reported trust in AI. Cai et al. \cite{cai_effects_2019} explored two kinds of example-based explanations in the sketch recognition domain, normative explanations and comparative explanations, and were deployed on the QuickDraw platform. Results from 1150 users showed that normative explanations helped improve system comprehension and perceived system capability, while comparative explanations revealed limitations. Accordingly, examples can serve as an effective explanatory tool, but different types of examples may have different advantages and disadvantages. Cecil’s research \cite{cecil_explainability_2024} further expands this discussion by exploring how individuals interact with AI-generated advice in personnel selection tasks. Across five experiments, they found that incorrect AI advice negatively affected performance due to overreliance, and that explainability methods like saliency map and charts had limited effects on mitigating this issue, highlighting the complexity of human-AI interaction in high-stakes contexts like HR management.  Leichtmann’s study \cite{leichtmann_effects_2023} explored both educational interventions and different forms of explainability such as attribution-based and example-based explanations, in the context of a mushroom-picking task. The findings suggest that while explainability and education about AI systems can influence task performance and user trust, domain-specific knowledge also plays a crucial role in shaping user outcomes. Similarly, Zhang et al. \cite{zhang_effect_2020} highlighted the importance of balancing human trust in AI with leveraging human expertise. Their study suggested that confidence scores as an explanation feature could help calibrate trust levels and improve collaboration between humans and AI, yet they also raised concerns about the limitations of local explanations, such as those provided by LIME, in decision-making processes.

Several studies have aimed to compare the effectiveness of global and local XAI methods in AI-assisted decision-making. Ahn et al. \cite{ahn_impact_2024} further examined the effects of explainability by comparing SHAP for global explanations and LIME for local explanations. Their findings revealed that, contrary to common expectations, interpretability alone did not lead to significant increases in trust, while outcome feedback had a stronger and more reliable impact on both trust and task performance. Similarly, Wang et al. \cite{wang_are_2021} presented an evaluation of established xAI methods in AI-assisted decision making, comparing their effectiveness in improving people's understanding of the AI model, helping people recognize model uncertainty, and supporting calibrated trust in the model. The study includes experiments on two types of decision-making tasks, recidivism prediction and forest cover prediction. They evaluated global methods, such as feature contribution and feature importance, and local methods, including counterfactual explanations and nearest neighbor example-based explanations. The results showed that the effectiveness of xAI methods largely depends on people's level of domain expertise and provide insights into designing and selecting effective xAI methods to better support human decision making. In another study \cite{alam_examining_2021}, participants interacted with a simulated AI system that initially gives the most likely, but incorrect diagnosis, but later changes the diagnosis to the correct one. Upon completion of further testing, they examine how a re-diagnosis event affects satisfaction and trust over time, and how different types of explanations (global and local) affect satisfaction, trust, and understanding of an AI system. In another experiment, they study how different forms of explanations (written, visual, and examples) in an AI diagnosis system affect patient satisfaction over time and find that richer explanations are the most effective at these critical points.

Lastly, the study closest to the goals of our paper is by Evans et al. \cite{evans_explainability_2022} that explores how digital pathologists interact with various xAI tools. This study qualitatively assesses pathologists' use of xAI techniques like saliency maps, concept attributions, prototypes, trust scores, and counterfactuals. These tools are designed to make AI recommendations in medical imaging clearer. While their study offers insight into pathologists’ preferences for easy-to-understand AI guidance, it stops short of measuring how these tools and different level of explainability affect pathologists' trust in practice and their subsequent reliance on them for decision-making.

\section{Methodology}
The goal of this study is to examine how varying levels of explainability in AI-based CDSS influence clinicians' trust and diagnostic performance in breast cancer detection. Additionally, the study explores how clinicians' demographic factors—such as age, gender, and familiarity with AI—affect their interaction with the AI system.

To achieve these objectives, we conducted a series of experiments where clinicians performed diagnostic tasks under different conditions of AI explainability. The study measured their diagnostic accuracy, decision time, and self-reported trust in the AI system. We also collected demographic data to analyze its impact on trust and performance.

The following sections describe the experimental design, experiment procedure, and measurements.

\subsection{Experimental Design}
\label{DoE}
The primary goal of this experiment is to assess the impact of varying levels of AI explainability on clinicians' trust and diagnostic accuracy in breast cancer diagnosis.


\begin{figure*}[h]
    \centering
    \includegraphics[width=0.8\linewidth]{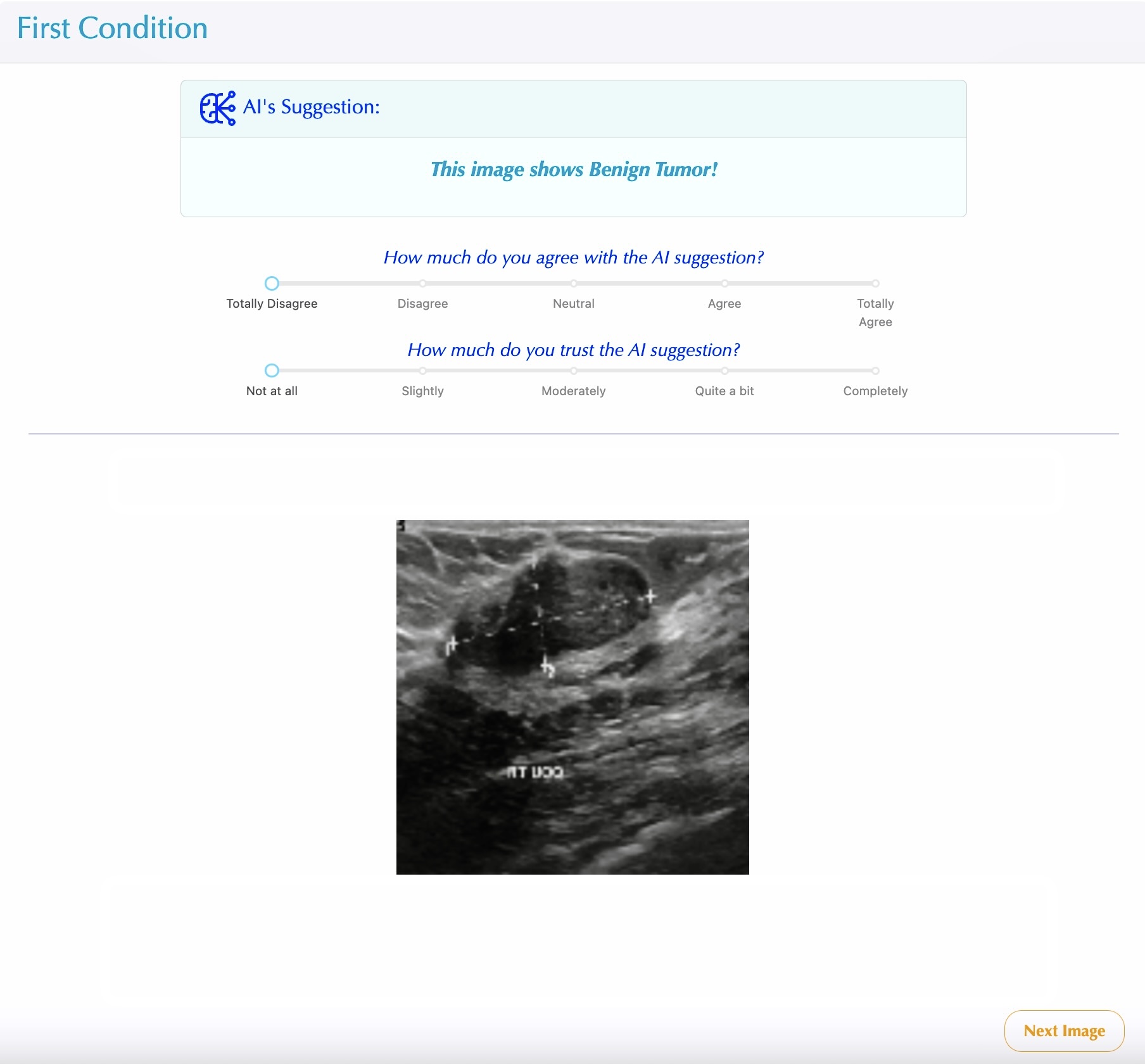}
    \caption{Experiment platform}
    \label{fig:first condition}
\end{figure*}

The target audience for this experiment is clinicians, including oncologists and radiologists, who traditionally process breast tissue scans to make cancer diagnosis decisions. To answer the research questions, we follow an interrupted time series (ITS) experiment design ~\cite{hartmann_interrupted_1980} where all participants use all versions of CDSSs (treated as multiple interventions) in a certain sequence. Initially, the CDSS presents diagnostic suggestions based on the analysis of breast cancer tissue images with a machine learning model without any accompanying explanations. These suggestions categorize the findings as 'healthy,' 'benign tumor,' or 'malignant tumor.' Subsequently, we introduce variations in the diagnostic process to understand how the level of explanation influences clinicians' trust and decision-making within the CDSS. All the clinical decisions recorded during the experiment are compared with the traditional process where the participants make diagnosis decisions in the absence of any decision support (treated as baseline). The experiments were conducted entirely online, without the need for direct oversight by the research team.

The ITS experiment process follows the interventions below in the given order. We designed the experiment such that the participants were exposed to decision support with an increasing level of explanations. \textit{\Cref{tab:conditions}} shows a detailed overview of the experimental conditions and their differences. Each condition involves diagnosing a series of ten breast tissue images, followed by post-experiment survey.

\begin{itemize}
    \item \textbf{\textit{Baseline (Stand-alone):}} Clinicians are not presented with any diagnostic suggestions and are asked to make diagnosis decisions on the breast cancer tissue images based on their own judgment.
    \item \textbf{\textit{Intervention I (Classification):}} Clinicians are presented with diagnostic suggestions (healthy, benign tumor, malignant tumor) without accompanying explanations.
    \item \textbf{\textit{Intervention II (Probability Distribution):}} Building upon the first, this intervention introduces probability estimates for each diagnostic class (healthy, benign tumor, malignant tumor).
    \item \textbf{\textit{Intervention III (Tumor Localization):}} In addition to the information provided in the second intervention, the CDSS advances by estimating the precise location of the tumor within the breast tissue images.
    \item \textbf{\textit{Intervention IV (Enhanced Tumor Localization with Confidence Levels):}} Compared to the third intervention, clinicians in this scenario receive tumor location information including both low and high confidence estimates upon tumor detection.
\end{itemize}

\begin{table*}

     \centering
     \caption{The experimental conditions involving AI suggestions and their descriptions}
     \begin{tabular}{ c  p{6cm} }
     \toprule
      AI Suggestion & Description \\ 
    \cmidrule(r){1-1}\cmidrule(l){2-2}
     \raisebox{-\totalheight}{\includegraphics[scale = 0.184]{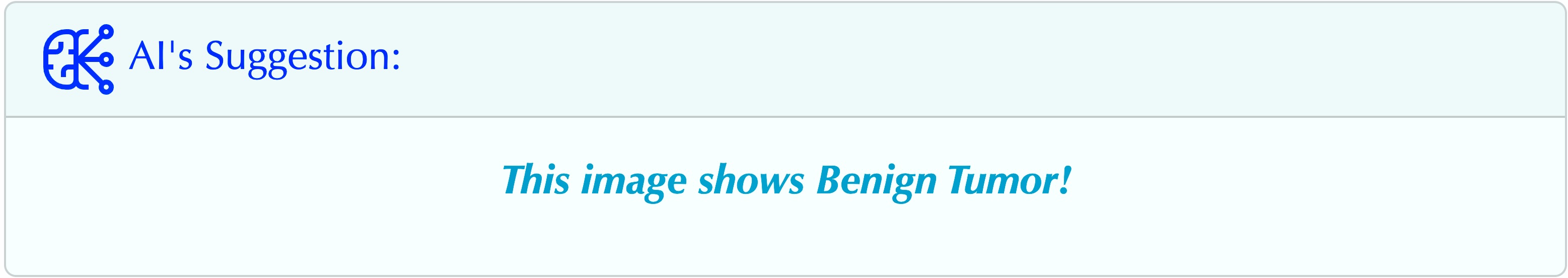}}
      & 
    
      The AI system provides diagnosis suggestions without any accompanying explanations, categorizing them into three distinct categories: healthy, benign tumor, and malignant tumor.
    
      \\ \bottomrule
      \raisebox{-\totalheight}{\includegraphics[scale= 0.266]{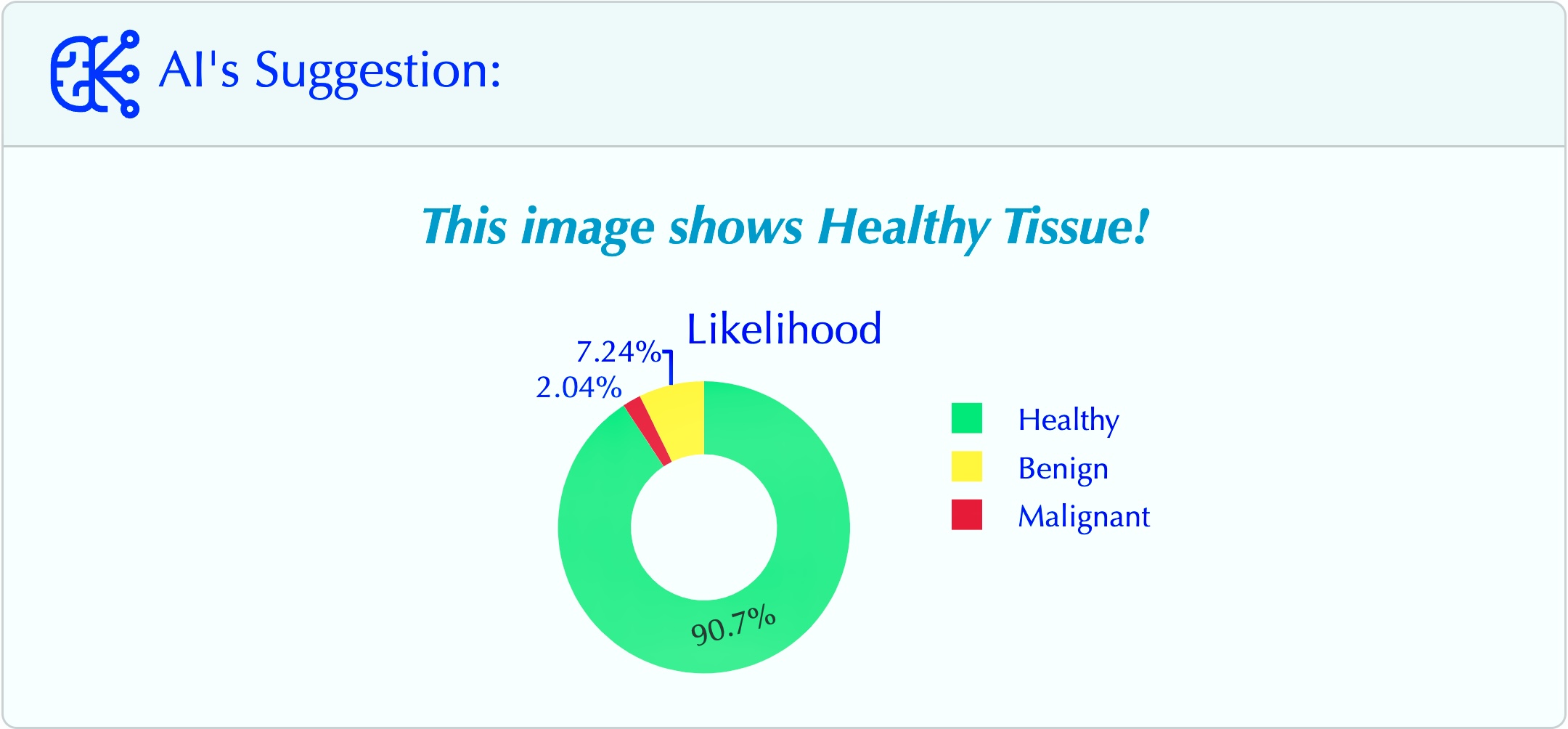}}
      & 
      The AI system presents diagnosis suggestions in three distinct categories—healthy, benign tumor, and malignant tumor—accompanied by corresponding confidence scores for each category.
      \\ \bottomrule
      \raisebox{-\totalheight}{\includegraphics[scale= 0.24]{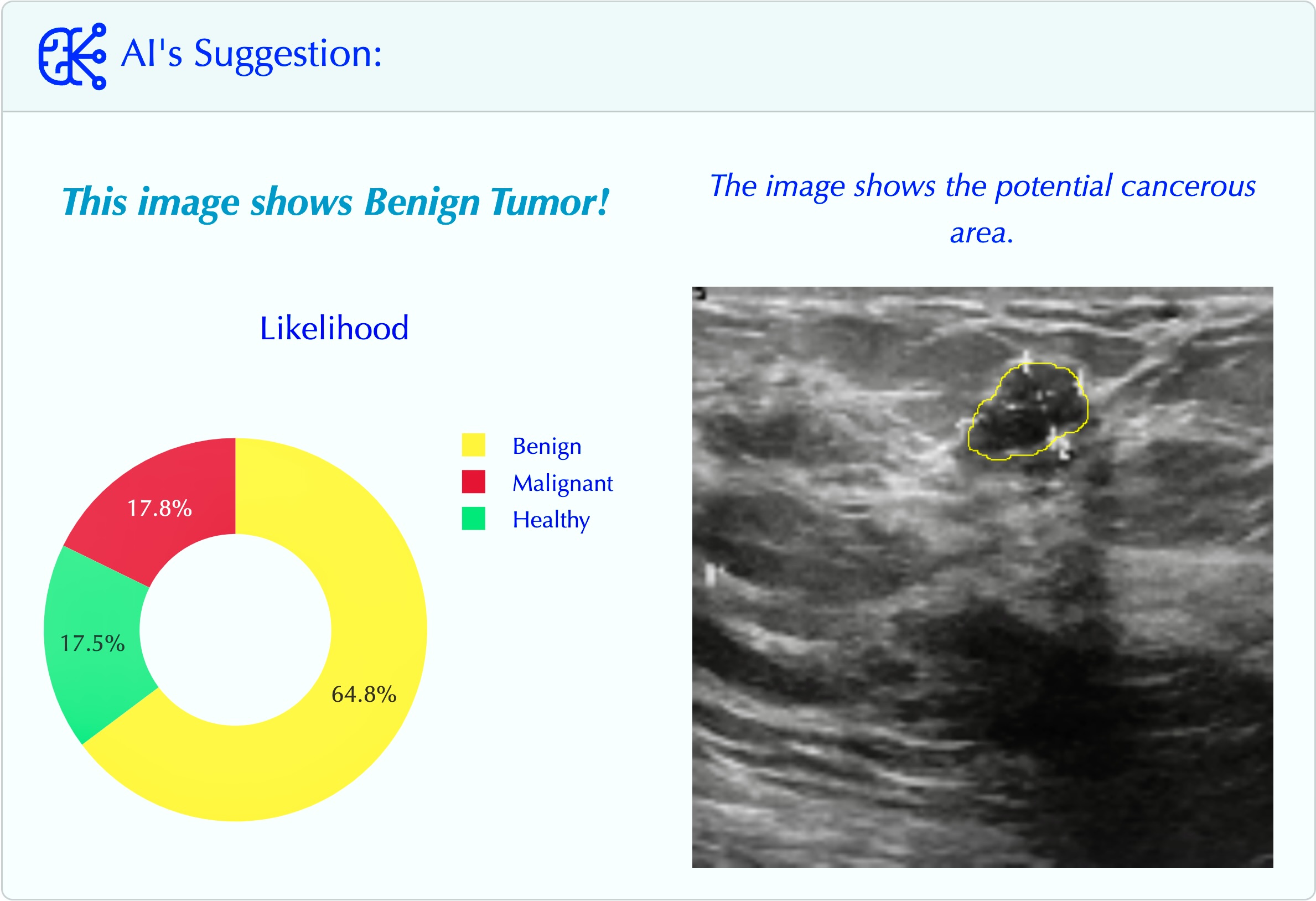}}
      & 
    
      The AI system provides diagnosis suggestions categorized into three distinct categories—healthy, benign tumor, and malignant tumor. Additionally, it offers corresponding probability scores for each category and estimates the location of the tumor, if detected.
    
      \\ \bottomrule
      \raisebox{-\totalheight}{\includegraphics[scale= 0.24]{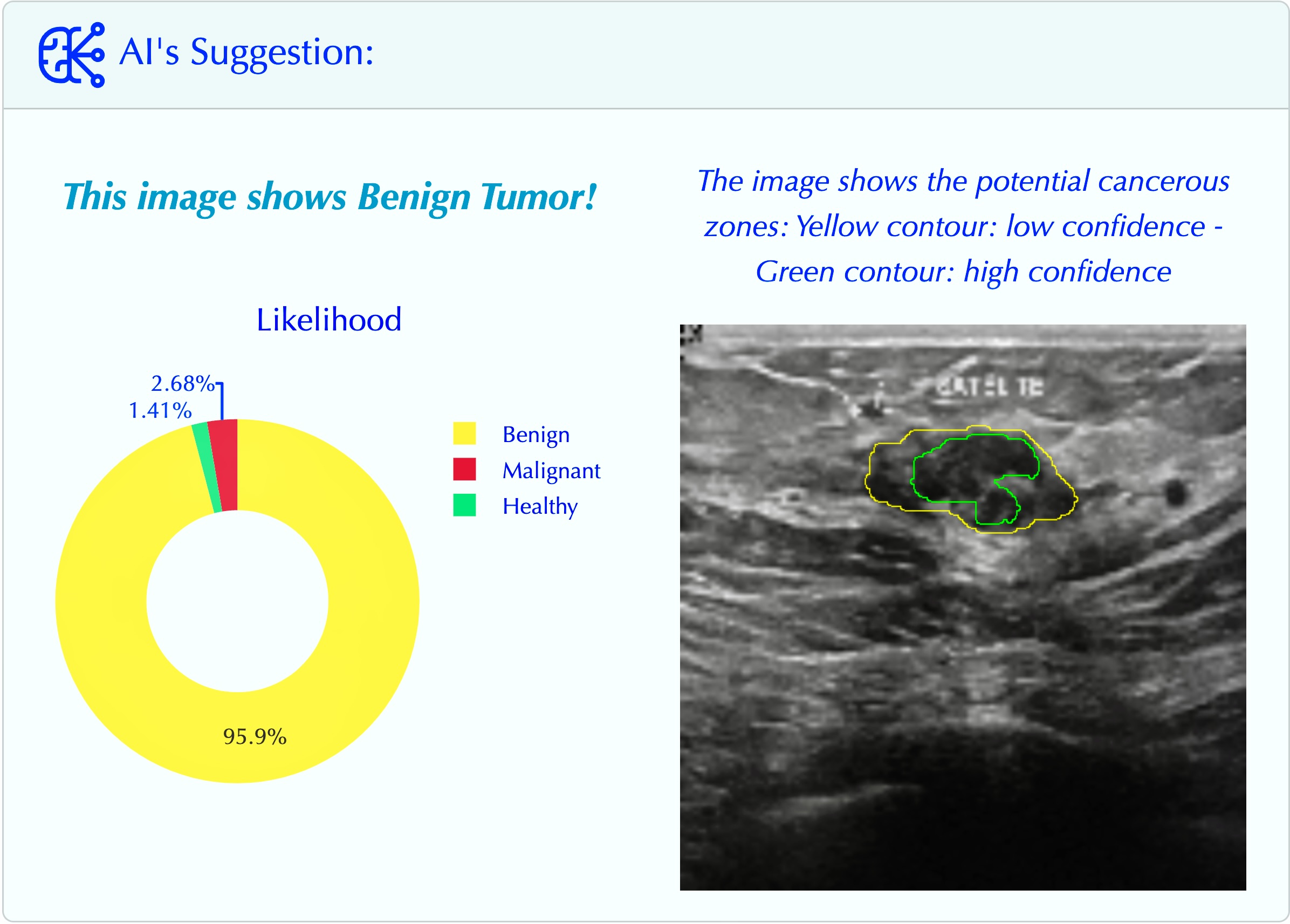}}
      & 
      The AI system presents diagnosis suggestions in three categories—healthy, benign tumor, and malignant tumor—along with associated confidence scores. Furthermore, it estimates the location of the tumor, providing both low confidence and high confidence if a tumor is detected.
      \\ \bottomrule
      \end{tabular}
      \label{tab:conditions}
\end{table*}

The AI system behind the CDSS builds upon our previous work~\cite{rezaeian_architecture_2024}. This system integrates a U-Net architecture for image segmentation with a Convolutional Neural Network (CNN) for cancer prediction \cite{ronneberger_u-net_2015}. The U-Net architecture was designed to depict the boundaries of cancerous areas in breast tissue, helping to localize potential tumors. The CNN then classifies the images into three categories: healthy, benign, and malignant. The system was trained on a publicly available breast cancer dataset \cite{al-dhabyani_dataset_2020} consisting of 780 ultrasound images, achieving an 81\% diagnostic accuracy. This accuracy result suggests that the tool may be useful for decision support if users exercise some healthy skepticism when following the recommendations provided by this tool.


\subsection{Experiment Procedure}
\label{procedure}

\subsubsection{Web Application}
The experiment platform we developed based on the Python-based Dash framework allows collecting quantitative data representing the clinical users' interaction and collaboration with an AI recommender system, coupled with both pre-and post-experiment surveys, all integrated into a single, user-friendly web application. This platform enables us to gain a comprehensive understanding of how medical professionals engage with AI-based CDSS in a virtual environment. The experiment process, starting with the submission of consent forms and going down to data collection, takes place online within this integrated web application.

\subsubsection{Recruitment}

We recruited clinicians aged 18 and older, including oncologists and radiologists experienced in interpreting breast tissue scans for cancer diagnosis. All participants were fluent in English. Recruitment efforts focused on medical associations, social media platforms, and professional networks, primarily within the USA.

\subsubsection{Tutorial Video and Preparation}

Interested participants contacted us via email to express their interest in the study. We sent them both a comprehensive video tutorial and the access link to the experiment via email. This tutorial explains the goals of the study, how to use the web application, and the tasks they need to complete. We emailed the link exclusively to those who showed interest, ensuring the authenticity of participants.

\subsubsection{Consent Form and Pre-Experiment Survey}

Participants signed a consent form electronically through the web app, followed by a pre-experiment survey that collected their demographic information and baseline data on their experience and comfort level with AI and clinical decision support systems in general. The survey collects the following information:

\begin{itemize}
    \item{\bfseries\textit{Demographic Information:}} This section asks for the participants' age, gender, and race/ethnicity.
    \item{\bfseries\textit{Professional Information:}} Participants also provide information regarding their current job, years of experience, and familiarity with AI systems.
    \item{\bfseries\textit{Experience with AI in Oncology or Radiology:}} This section checks if participants have used AI systems in their field before.
    \item{\bfseries\textit{Trust in AI Systems:}} This part measures how much participants trust AI systems in oncology or radiology.
    \item{\bfseries\textit{Opinions on AI in Medicine:}} Participants share their thoughts on AI's future in medicine, its challenges in clinical settings, and its usefulness in their jobs.
\end{itemize}

\subsubsection{Experimental Sessions}

After the pre-experiment survey, participants engaged in a series of experimental sessions discussed in \textit{\Cref{DoE}}. While each experiment condition differs in terms of the amount of information presented to participants, in all these scenarios, participants were presented with a series of ten mammogram images one at a time. Their primary task in the baseline condition was to examine each image and provide one of three diagnoses: Healthy, Benign, or Malignant. This baseline condition aimed to evaluate their diagnostic performance without the assistance of AI. 

\begin{figure}[t]
    \centering
    \includegraphics[width=0.8\linewidth]{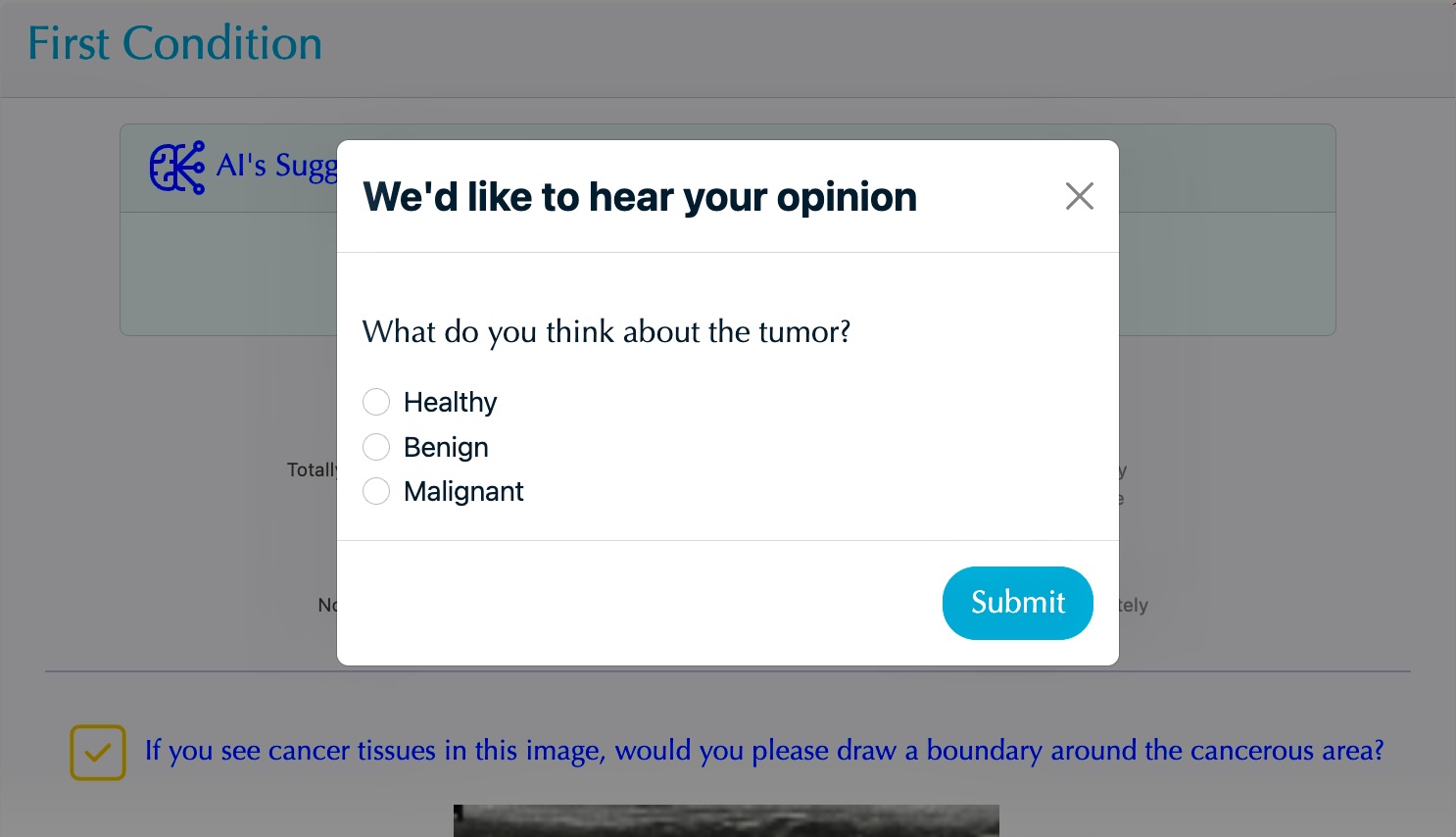}
    \caption{Interactive evaluation: Participants providing their input when in disagreement with AI during the diagnostic process}
    \label{fig: modal}
\end{figure}

The conditions that involve CDSS presented the AI suggestions and their corresponding descriptions. In those conditions, participants were also instructed to assess their agreement with and trust the AI's suggestions for each image they viewed. They rated their agreement and trust levels on a Likert scale using sliders (see \textit{\Cref{fig:first condition}}). If their agreement level fell below neutral, they were asked to make their own decision, as shown in \textit{\Cref{fig: modal}}. This approach allowed us to capture their true decision for images where they did not agree with the AI. 

\subsubsection{Post-Experiment Survey}

Upon completion of all experimental sessions, participants proceeded to the post-experiment survey. The questions in this survey are designed to gather specific insights into their experience and assess the differences in each condition in terms of their understanding, trust, and workload. Participants were asked to respond to the following questions using a 5-point Likert scale:

\begin{enumerate}[label=Q\arabic*:]
\setlength{\itemsep}{-0.5em}
    \item I believe that the AI answers were accurate.
    \item How mentally demanding was the task?
    \item How stressed, and annoyed were you?
    \item I found the AI's suggested breast cancer classification (Healthy, Benign, or Malignant) to be intuitively understandable.
    \item I could access a great deal of information which explained how the AI system worked.
    \item The AI suggestions helped me to decide whether I can trust the generated suggestions.
    \item I trusted the AI suggestions in the diagnosis of breast cancer.
\end{enumerate}




\subsection{Measurements}
In this study, we assess a variety of measures related to trust, performance, and user interactions in AI-based breast cancer diagnosis. These measures are categorized into self-reported and behavioral measures following Kohn et al. \cite{kohn_measurement_2021}. 

\subsubsection{Self-Reported Measures}
These metrics capture participants' perceptions and beliefs about the AI system through surveys. These measures are often used to capture subjective aspects by asking participants to rate their experiences on Likert scales.

\begin{itemize}
    \item {\bfseries\textit{Trust: }}We assessed participants' trust in AI systems using a 5-point Likert scale from 0 (No trust at all) to 5 (Complete trust). During the experiment, participants rated their trust in AI suggestions for each image on a similar 5-point scale, responding to the question, ``How much do you trust the AI system?'', providing a finer measurement.
    \item {\bfseries\textit{Familiarity: }}In the demographic survey, we assessed participants' familiarity with AI systems using a 5-point Likert scale. Participants were asked, ``How familiar are you with Artificial Intelligence systems?'' with response options ranging from ``Not at all'' to ``Completely familiar''.
    \item {\bfseries\textit{Understandability: }}After each intervention in the post-experiment survey, participants were asked to evaluate the understandability of the AI suggestions. The statement, ``I found the AI's suggested breast cancer classification to be intuitively understandable'', was rated on a 5-point Likert scale, ranging from ``Strongly disagree'' to ``Strongly agree''.
    \item {\bfseries\textit{Perceived Accuracy: }}After each intervention, in the post-experiment survey, we assessed participants' perceptions of the AI's accuracy by asking them to respond to the statement, ``I believe that the AI answers were accurate.'' Participants rated the perceived accuracy of the AI suggestions on a scale ranging from ``Not accurate'' to ``Completely accurate.''
\end{itemize}

\subsubsection{Behavioral Measures}
These metrics assess observable actions that indicate a participant's trust in the AI system.
\begin{itemize}
    \item {\bfseries\textit{Performance: }}Participants' performance was assessed based on their decisions about each image during the experiment. To determine their performance, we considered two factors. First, if the participant's agreement level with the AI suggestion was higher than neutral, we considered their decision to align with the AI's suggestion. For cases where the agreement level was below neutral, we asked participants to independently indicate their decision by choosing among ``Healthy'', ``Benign'', or ``Malignant'', as illustrated in \textit{\Cref{fig: modal}}. We calculated their performance for each intervention by comparing their decisions with the ground truth over ten images as a percentage, reflecting the accuracy of their decisions relative to the correct diagnosis.
    
    \item {\bfseries\textit{Agreement Rate: }}During the experiment, participants rated their agreement with AI suggestions for each image on a 5-point scale, responding to the question, ``How much do you agree with AI suggestion?''
    
    \item {\bfseries\textit{Decision Time: }}We tracked the duration of participants' decision-making process for each image throughout the experiment. 
    
    
\end{itemize}

\begin{table}[b]
\centering
\caption{Study participant characteristics (n=28)}

\resizebox{0.75\columnwidth}{!}{ 
\Large
\begin{tabular}{llc}
                  
\textbf{Characteristics}    & \textbf{} & \textbf{} \\
\hline
Age Bracket (years)          & $25-35$   & $5$  \\
                      & $35-45$   &  $15$ \\
                      & $45-55$   &  $3$ \\
                      & $55+$ & $5$\\
\hline
Gender(n)            & Female   & $11$  \\
                    & Male     & $17$ \\
\hline
Ethnicity(n)         & White   & $19$\\
                    & Asian   & $4$\\
                    & Other  & $2$\\
                    & Middle Eastern, North African   & $2$\\
                    & Hispanic, Latino, Spanish    & $1$\\
                   
\hline
Medical Role(n)     & Radiologist    & $17$\\
                    & Oncologist    & $5$\\
                    & Other   & $6$\\

\end{tabular}}
\label{tab:Demographics}
\end{table}
\section{Results}

\subsection{Sample Demographics and Professional Background}
\begin{figure}
    \centering
    \includegraphics[width=1\linewidth]{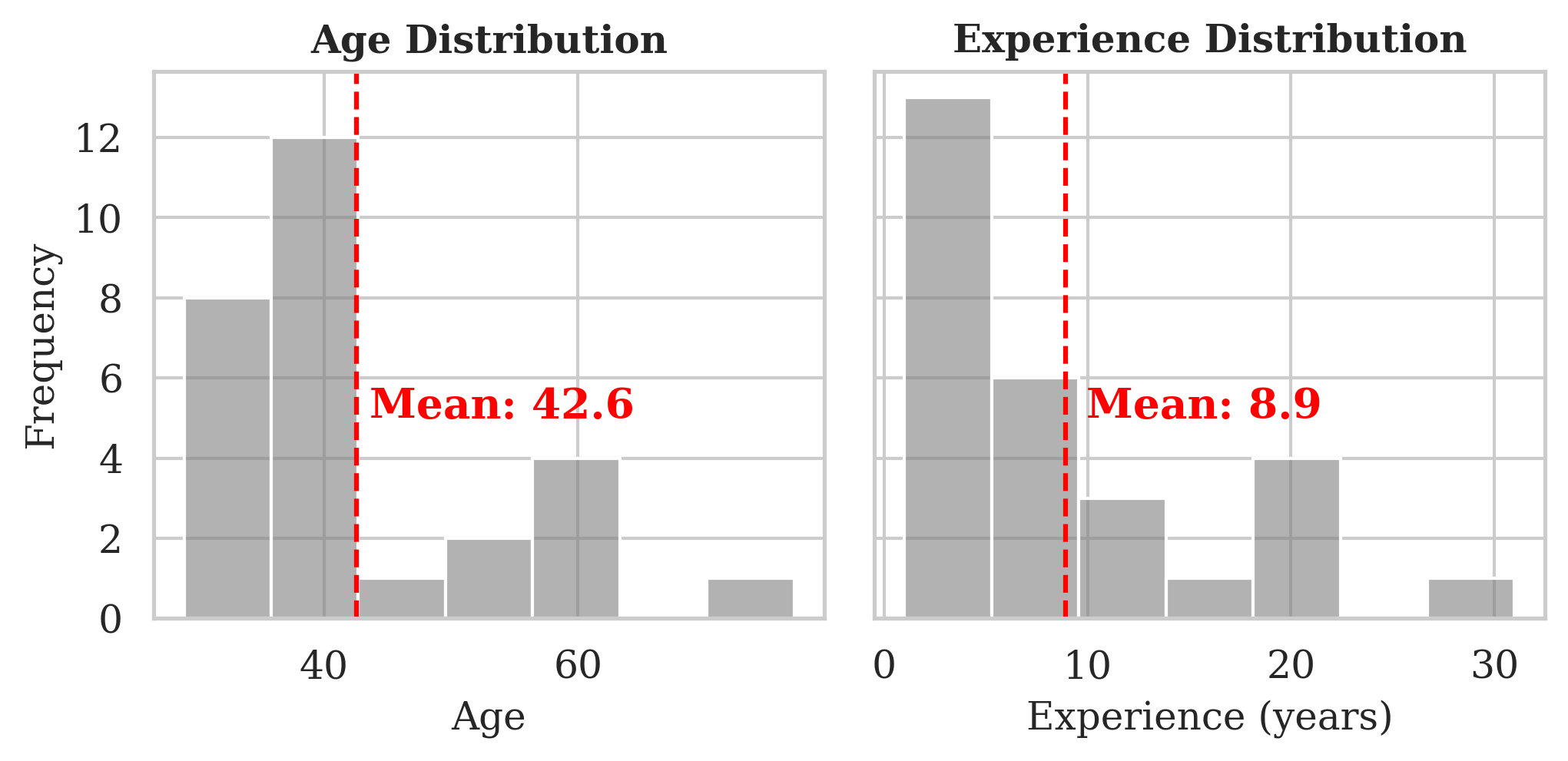}
    \caption{Age and experience distribution of participants}
    \label{fig:age-experience}
\end{figure}

The sample consisted of 28 participants (\textit{\Cref{tab:Demographics}}), including 17 males ($\approx60\%$) and 11 females ($\approx40\%$). The participants' age ranged from 29 to 77, with a mean age of 42.6 years ($SD = 12.1$). 18\% of the participants were oncologists, 60\% were radiologists, and the remaining 22\% held other medical roles related to breast cancer diagnosis such as breast surgeons. The experience in the profession ranged from 1 to 31 years, with a mean of 8.9 years ($SD = 7.3$), as seen in \textit{\Cref{fig:age-experience}}. Correlation analysis revealed that age and years of experience were highly correlated ($r=0.76$). Additionally, around 65\% of the participants reported having prior experience using AI systems in their job to assist with diagnosing breast cancer. A chi-square test further indicated that, among the participants, radiologists were significantly more likely to have used AI in their job $(\chi^2 = 6.2, P = 0.04)$.

\begin{table}[hb]
\centering
\caption{Chi square test results: Effect of age and gender on categorical measures}
\resizebox{0.95\columnwidth}{!}{ 
\Large
\begin{tabular}{lcccccccc}
\hline
                     & \multicolumn{2}{c}{\textbf{Familiarity}}  & \multicolumn{2}{c}{\textbf{Understandability}} & \multicolumn{2}{c}{\textbf{Perceived Accuracy}} \\
\hline
\textbf{Variable}    & \textbf{$\chi^2$} & \textbf{$p$} & \textbf{$\chi^2$} & \textbf{$p$} & \textbf{$\chi^2$} & \textbf{$p$} \\
\hline
Gender            &24.8    & \textbf{5.52e-05}  & 6.95      & 0.14     & 9.38   & 0.05 \\
Age Bracket       & 49.5   & \textbf{1.73e-08}  & 29.83      & \textbf{2.95e-03}    & 32.09  & \textbf{1.33e-03}\\
Experience Bracket       & 18.2   & \textbf{0.02}  &  17.6     & \textbf{0.02}    & 14.1 &0.08\\

\hline
\end{tabular}
}
\label{tab:Chi-Square}
\end{table}

\begin{table}[htp]
\centering
\vspace{-1em}
\caption{ANOVA results: Effect of age and gender on quantitative measures}
\resizebox{0.95\columnwidth}{!}{ 
\Large
\begin{tabular}{lcccccccc}
\hline
                     & \multicolumn{2}{c}{\textbf{Performance}} & \multicolumn{2}{c}{\textbf{Diagnosis Duration}}& \multicolumn{2}{c}{\textbf{Trust}}& \multicolumn{2}{c}{\textbf{Agreement}}  \\
\hline
\textbf{Variable}    & \textbf{$SS$} & \textbf{$p$} & \textbf{$SS$} & \textbf{$p$}& \textbf{$SS$} & \textbf{$p$} & \textbf{$SS$} & \textbf{$p$} \\
\hline
Gender          & 0.01 & 0.53 & 0.49  & 0.11  &0.13& 0.56&0.06&0.65\\
Age Bracket     & 0.09 & 0.18 & 1.57  & \textbf{0.04}&0.28&0.86&1.15&0.30\\
Experience Bracket    & 0.03 & 0.47 & 0.02 &0.95 &1.23&0.21&0.52&0.43\\
\hline
\end{tabular}
}
\label{tab:Anova}
\end{table}

\begin{figure}
    \centering
    \includegraphics[width=1\linewidth]{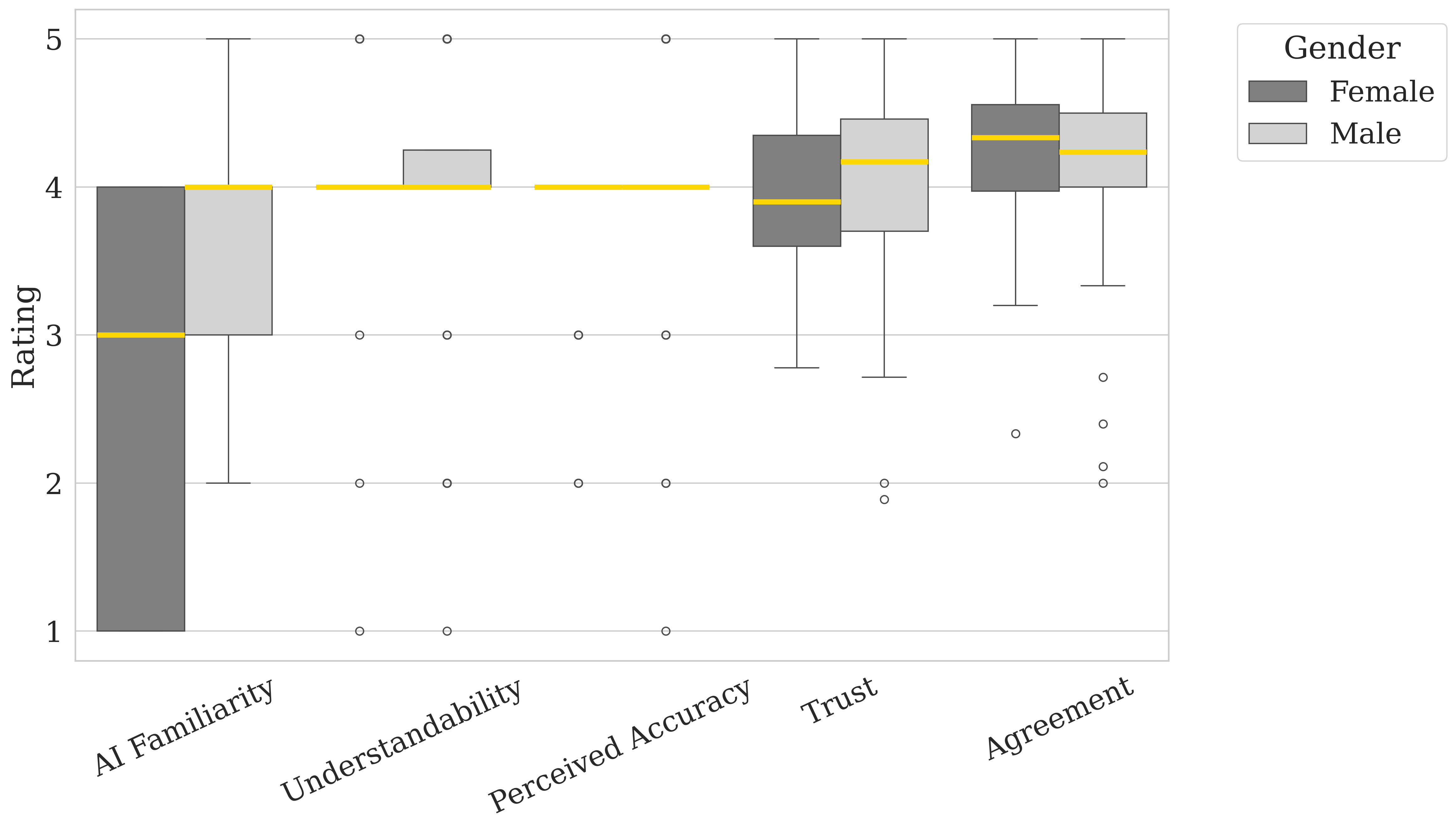}
    \caption{Distribution of AI-related measures by gender}
    \label{fig:Demo}
\end{figure}

A series of chi-square tests (see \textit{\Cref{tab:Chi-Square}}) revealed interesting relationships between key measures (familiarity, understandability, perceived accuracy, trust, and agreement) and participant demographics (gender, experience bracket and age bracket). 
In these results, gender was significantly associated with familiarity with AI systems ($\chi^2 = 24.8$, p = 5.52e-05). Male participants reported significantly higher familiarity with AI systems compared to female participants as shown in \textit{\Cref{fig:Demo}}. 

\begin{figure}
    \centering
    \includegraphics[width=1\linewidth]{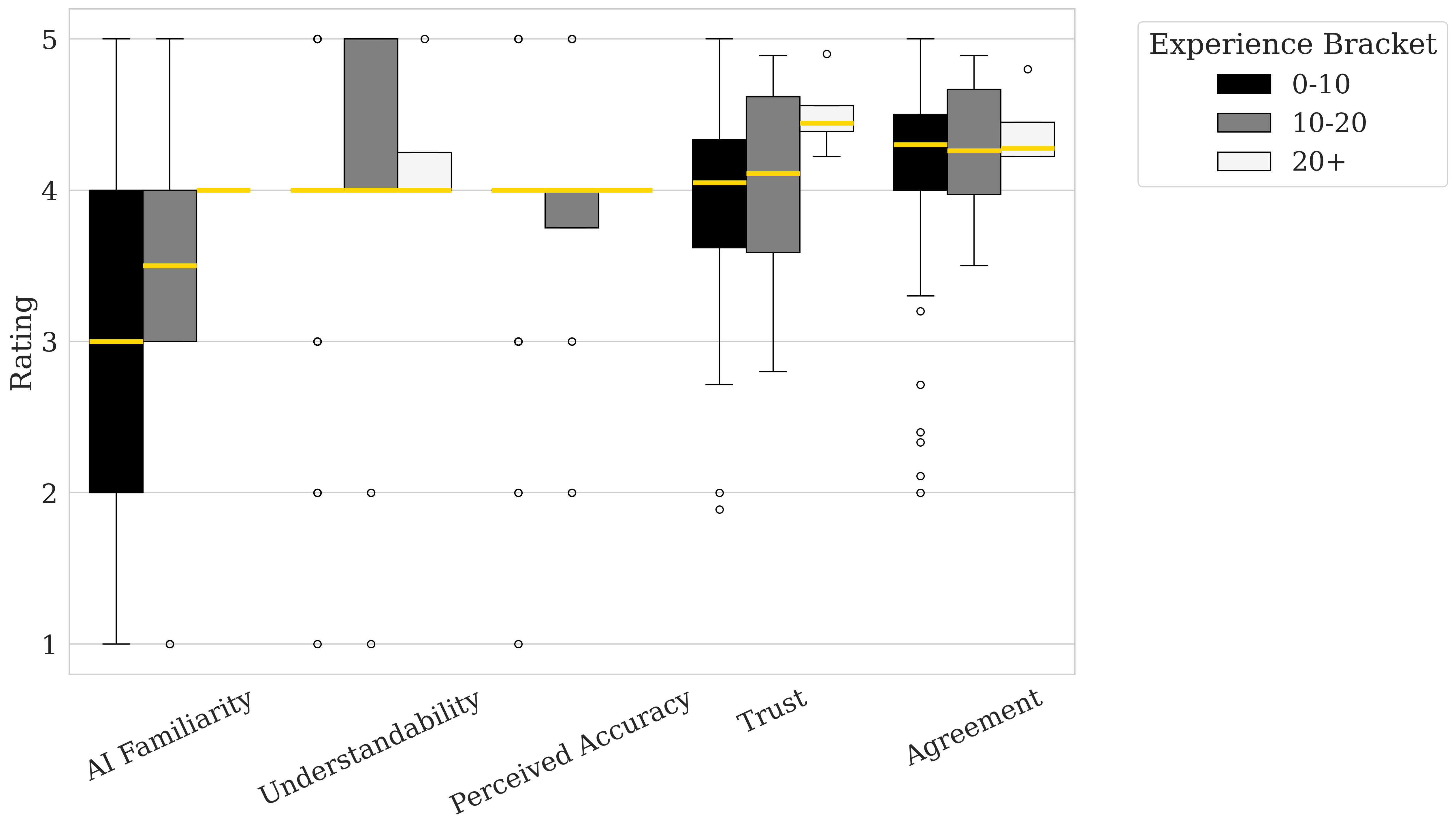}
    \caption{Distribution of AI-related measures by experience}
    \label{fig:Demo2}
\end{figure}

Furthermore, the analysis revealed significant associations between participants' experience bracket and key measures, as seen in \textit{\Cref{fig:Demo2}}. Participants with higher levels of experience were significantly more familiar with AI systems ($\chi^2 = 18.2, p = 0.02$), suggesting that those with more exposure over longer periods of time in clinical settings had developed a deeper understanding of AI technologies. Further analysis of understandability showed that the participants with 10-20 years of experience found the AI systems significantly more understandable than others ($\chi^2 = 17.6, p = 0.02$).

\begin{figure}
    \centering
    \includegraphics[width=1\linewidth]{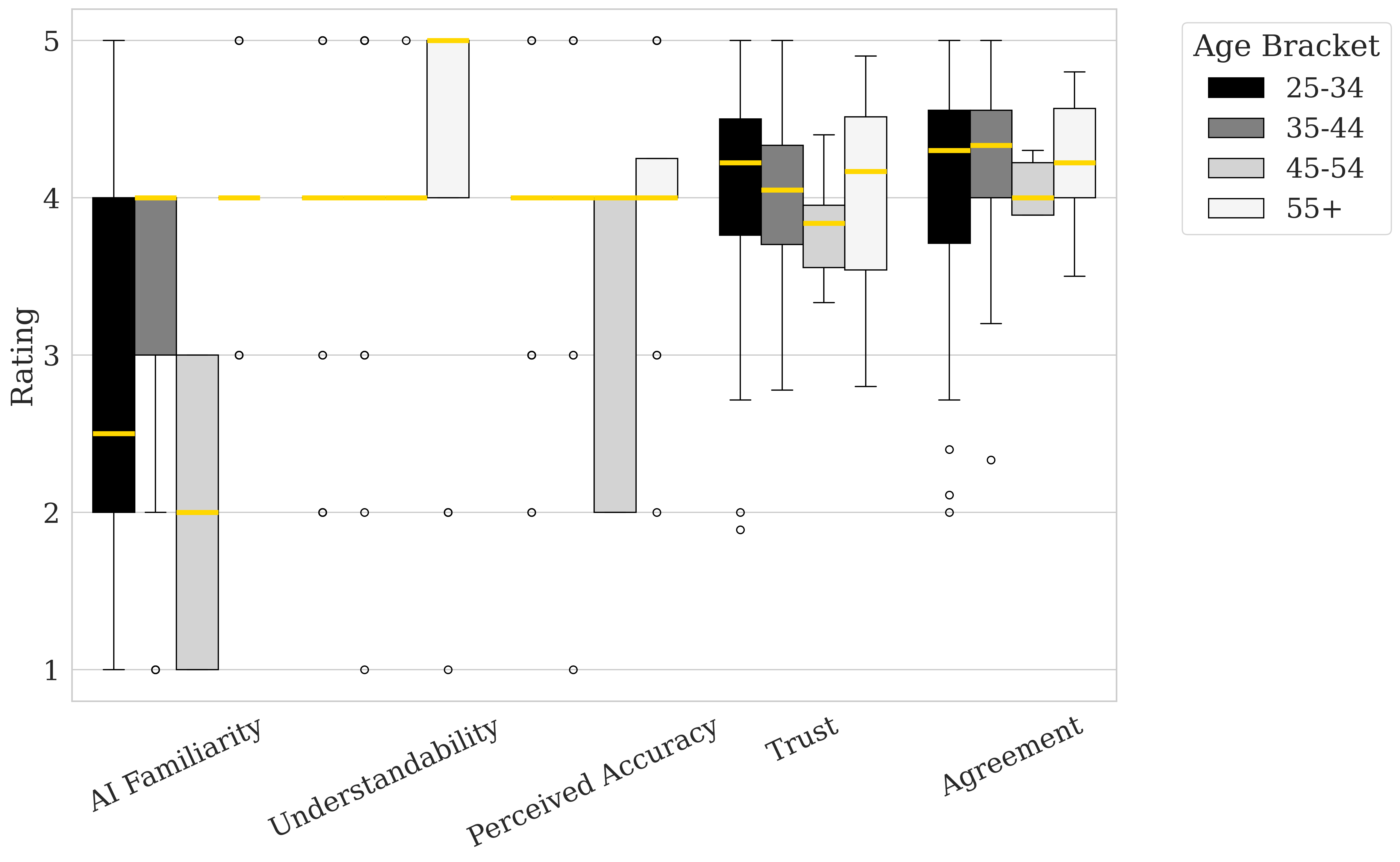}
    \caption{Distribution of AI-related measures by age}
    \label{fig:Demo1}
\end{figure}

The same results revealed significant relationships between age brackets and several key measures (\textit{\Cref{fig:Demo1}}). Regarding AI familiarity ($\chi^2 = 49.5, p = 1.73e-08$), older participants reported significantly higher familiarity with AI systems, which may be attributed to greater professional experience and more opportunities to engage with AI in clinical settings. Additionally, understandability ($\chi^2 = 30.7$, p = 0.014) was significantly associated with age, as participants in older age brackets found AI systems more understandable compared to younger participants. Perceived accuracy ($\chi^2 = 28.19$, p = 0.03) also showed a significant relationship with age, with older participants ($55+$) rating the AI as more accurate.

\begin{figure}
    \centering
    \includegraphics[width=1\linewidth]{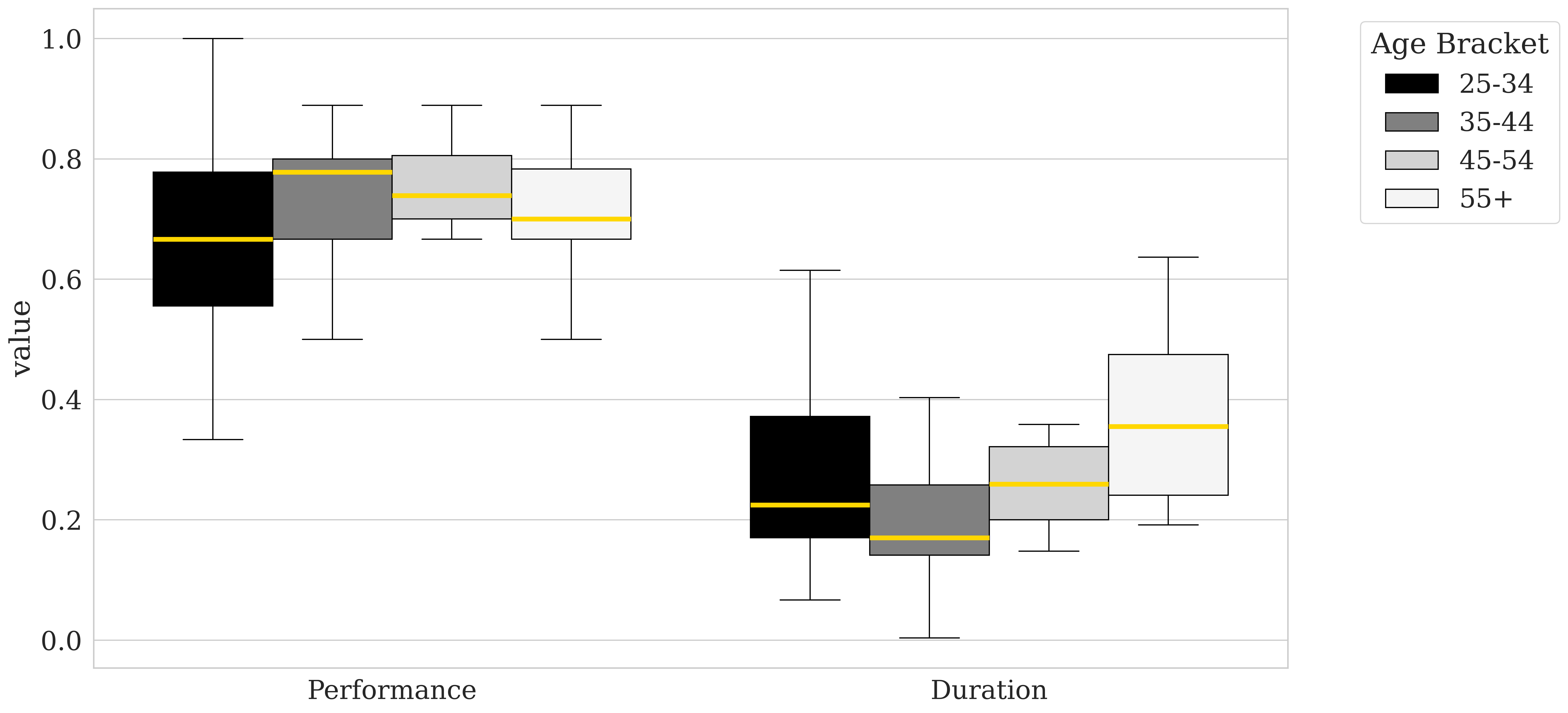}
    \caption{Distribution of participant performance and diagnosis duration by age bracket.}
    \label{fig:boxplot}
\end{figure}

The ANOVA analysis (see \textit{\Cref{tab:Anova}}) showed no significant effect of gender on performance $(p = 0.53)$, diagnosis duration $(p = 0.11)$, trust $(p = 0.56)$, and agreement $(p = 0.65)$ indicating no major differences between males and females in these measures. Similarly, age bracket was not significantly associated with performance $(p = 0.18)$, trust $(p = 0.86)$, and agreement $(p = 0.30)$ but was significantly associated with diagnosis duration $(p = 0.04)$. As \textit{\Cref{fig:boxplot}} shows, older participants, particularly those who were 55+ tend to spend more time on diagnosis. 

\subsection{Self-Reported \& Behavioral Measures Across Interventions}

\begin{figure*}[tp]
    \centering
    \includegraphics[width=\textwidth]{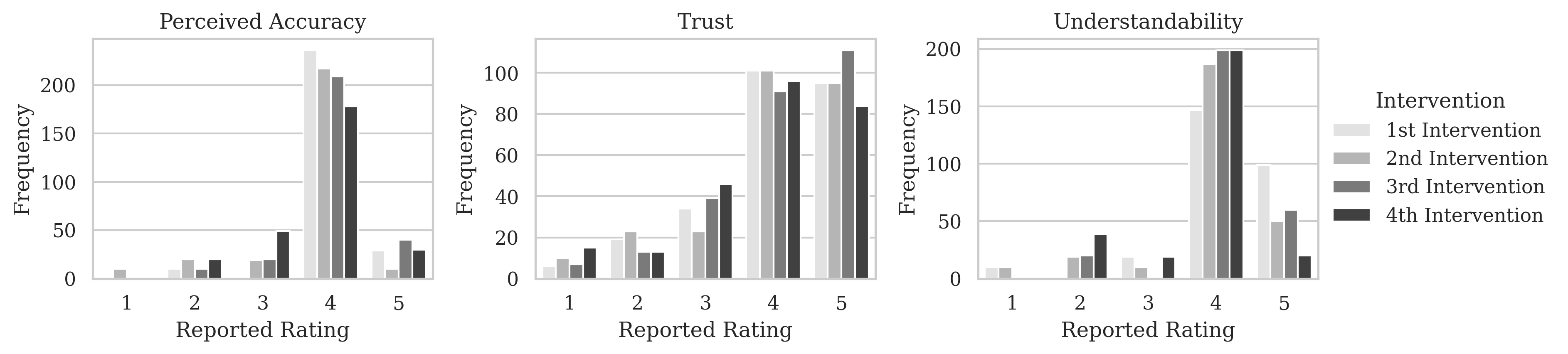}
    \caption{Distribution of reported values for perceived accuracy, understandability, and trust across Interventions.}
    \label{fig:image3}
\end{figure*}

Our analysis with a mixed-effects model in \textit{\Cref{tab:mixed-results1}} along with the distributions presented in \textit{\Cref{fig:image3}} and \textit{\Cref{fig:boxacc}} provide insights into how varying levels of AI explainability impact the self-reported and behavioral measures we assessed during the experiment.

\subsubsection*{Trust:}
The mixed-effects model revealed no statistically significant effects of the interventions on trust. Participants reported a minimal negative effect on trust in the 2nd and 4th interventions compared to the 1st, as indicated by the coefficient of -0.049 ($p = 0.543$) and -0.145 ($p = 0.073$), respectively, while the 3rd intervention showed a minimal positive effect with a coefficient of 0.059, ($p = 0.187$), with none of them being statistical significant. 
These findings suggest that increasing the level of AI explainability does not always enhance trust.

However, when considering participants' baseline trust in AI systems (prior to any interventions) as a reference point in the mixed-effects model, the analysis revealed that overall trust increased after interacting with the system, regardless of the CDSS type providing the explanations. The coefficients for each intervention are 0.888 ($p = 0.000$), 0.836 ($p = 0.000$), 0.944 ($p = 0.000$) and 0.750 ($p = 0.000$), respectively. Notably, the 3rd intervention led to the highest increase in trust with a coefficient of 0.944.


\subsubsection*{Understandability:}
The 4th intervention resulted in significantly reduced understandability of the AI system compared to the 1st with a coefficient of -0.464 ($p = 0.013$). This highlights a potential trade-off between providing detailed explanations and maintaining clarity. These suggest that while moderate levels of explainability may have little impact on users' understanding, providing too much information can lead to confusion and lower overall understandability.

\begin{table}[htp]
\centering
\caption{Mixed linear model results: Effect of AI explainability on self-reported \& behavioral measures}
\resizebox{0.95\columnwidth}{!}{ 
\large
\begin{tabular}{lcccccc}
\hline
\textit{Self-reported}           & \multicolumn{2}{c}{\textbf{Trust}}  & \multicolumn{2}{c}{\textbf{Understandability}} & \multicolumn{2}{c}{\textbf{Perceived Accuracy}} \\
\hline
\textbf{Variable}    & \textbf{Coef.} & \textbf{$p$} & \textbf{Coef.} & \textbf{$p$} & \textbf{Coef.} & \textbf{$p$} \\
\hline
Intercept            & 3.031     & \textbf{0.000}      & 3.179      & \textbf{0.000}    & 3.036        &\textbf{0.000} \\
2nd Intervention     & -0.049    & 0.543               & -0.286     & 0.126             & -0.321       & \textbf{0.011} \\
3rd Intervention     & 0.059     & 0.465               & -0.107     &0.566              & -0.036       & 0.778 \\
4th Intervention     & -0.145    & 0.073               & -0.464     & \textbf{0.013}    & -0.250       &\textbf{0.048} \\
Group Var            & 0.294     &       -              & 0.201      &     -              &   0.233   & - \\
\hline
\end{tabular}
}
\begin{tabular}{lcccccc}
\hline
 & & & & & & \\
\hline
\end{tabular}
\resizebox{0.95\columnwidth}{!}{ 
\large
\begin{tabular}{lcccccc}
\hline
 \textit{Behavioral}      & \multicolumn{2}{c}{\textbf{Agreement}} &\multicolumn{2}{c}{\textbf{Diagnosis Duration}} &\multicolumn{2}{c}{\textbf{Performance}} \\
\hline
\textbf{Variable}    & \textbf{Coef.} & \textbf{$p$} & \textbf{Coef.} & \textbf{$p$} & \textbf{Coef.} & \textbf{$p$} \\
\hline
Intercept            & 3.224    & \textbf{0.000} & 0.260  & \textbf{0.000}   &0.736      & \textbf{0.000}\\
2nd Intervention     & -0.053   & 0.482          & 0.054  & 0.338            &0.005      & 0.835\\
3rd Intervention     & 0.035    & 0.640          & 0.009  & 0.869            &-0.073      & \textbf{0.005}\\
4th Intervention     & -0.149   & \textbf{0.048} & 0.144  & \textbf{0.011}   &-0.068      & \textbf{0.009}\\  
Group Var            & 0.24     &   -            & 0.094  &  -               & 0.008     &    -           \\
\hline
\end{tabular}
}

\label{tab:mixed-results1}
\vspace{-1em}
\end{table}

\subsubsection*{Perceived Accuracy:}
The results for perceived accuracy show that both the 2nd and 4th interventions lead to significantly lower perceived accuracy compared to the 1st with coefficient of -0.321, ($p = 0.011$), and -0.250 ($p = 0.048$), respectively. This suggests that while AI explanations are intended to improve user confidence in the system’s decisions, detailed information may undermine this confidence, leading to lower perceptions of accuracy.

\subsubsection*{Performance:}
\begin{figure}[htp]
    \centering
    \includegraphics[width=1\linewidth]{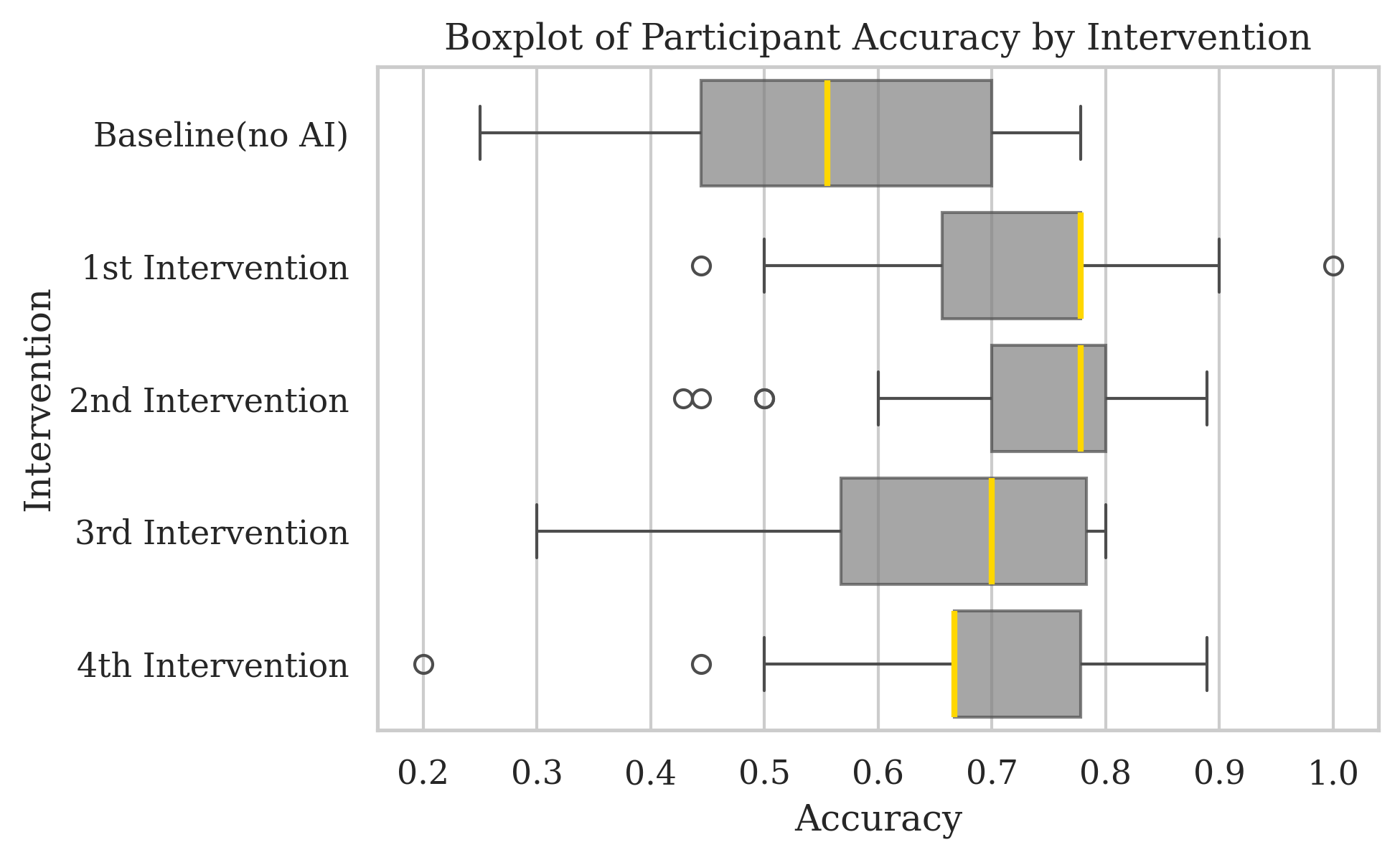}
    \caption{Participant performance over different interventions}
    \label{fig:boxacc}
\end{figure}

Results in \textit{\Cref{fig:boxacc}} shows that the accuracy rates when using AI (regardless of the intervention type) were generally higher compared to the baseline intervention without AI, all improvements being statistically significant. The accuracy rates across interventions demonstrate both improvements and declines in performance depending on the intervention. The medians in the 1st and 2nd interventions were the highest, indicating better overall performance in these AI-assisted interventions. Further analysis in \textit{\Cref{tab:mixed-results1}} shows that while the second intervention does not show any improvement compared to the 1st ($p=0.835$), the 3rd and 4th interventions show statistically significant reductions in diagnosis accuracy with coefficients of -0.073 ($p=0.005$) and -0.068 ($p=0.009$).

We also evaluated participants' performance in diagnosing images as healthy, benign, or malignant, with accuracy rates of 0.75, 0.68, and 0.68, respectively. The mixed-effects model revealed that accuracy the reduction in accuracy for benign and malignant diagnoses was significantly lower compared to healthy diagnoses.

\subsubsection*{Agreement Rate:}
As shown in \textit{\Cref{tab:mixed-results1}}, the mixed-effect model shows a significant difference in agreement levels between the 1st intervention (no explanation) and the 4th intervention (highest level of explainability), showing the agreement level in the 4th intervention was significantly lower than in the 1st intervention, with a coefficient of -0.149 $(p = 0.048)$. 

Further analysis reveals a strong positive correlation between participants' agreement with AI suggestions and their trust in the AI system, with a Spearman correlation coefficient of 0.85 $(p < 0.05)$. These findings suggest that the alignment between AI recommendations and participants' own decisions is critical in trust formation in AI-assisted decision-making.

\subsubsection*{Diagnosis Duration:}
When analyzing the effect of different interventions on the duration of decision-making, the mixed-effects model in \textit{\Cref{tab:mixed-results1}} demonstrates that the 4th intervention, which involved the highest level of explainability, led to longer decision times, compared to the 1st intervention. This suggests that more comprehensive explanations provided by the AI system may increase cognitive load and decision time.

\section{Discussion}


\subsection{Main Findings}
An interesting observation can be made about the role of gender in the results. While female participants reported lower familiarity with AI, there is no difference in trust or performance between male and female participants when using the AI-based CDSS. The same outcome was observed in our previous study using a sketch recognition game with 47 university students~\cite{lotfalian_saremi_trust_2024}. This finding further reinforces the notion that differences in self-reported familiarity between genders may not be a good indicator when designing AI systems for human use. Gender may not be a significant factor to consider when designing AI systems based on the lack of difference in behavioral measures. 

\begin{figure}[htp]
    \centering
    \includegraphics[width=1\linewidth]{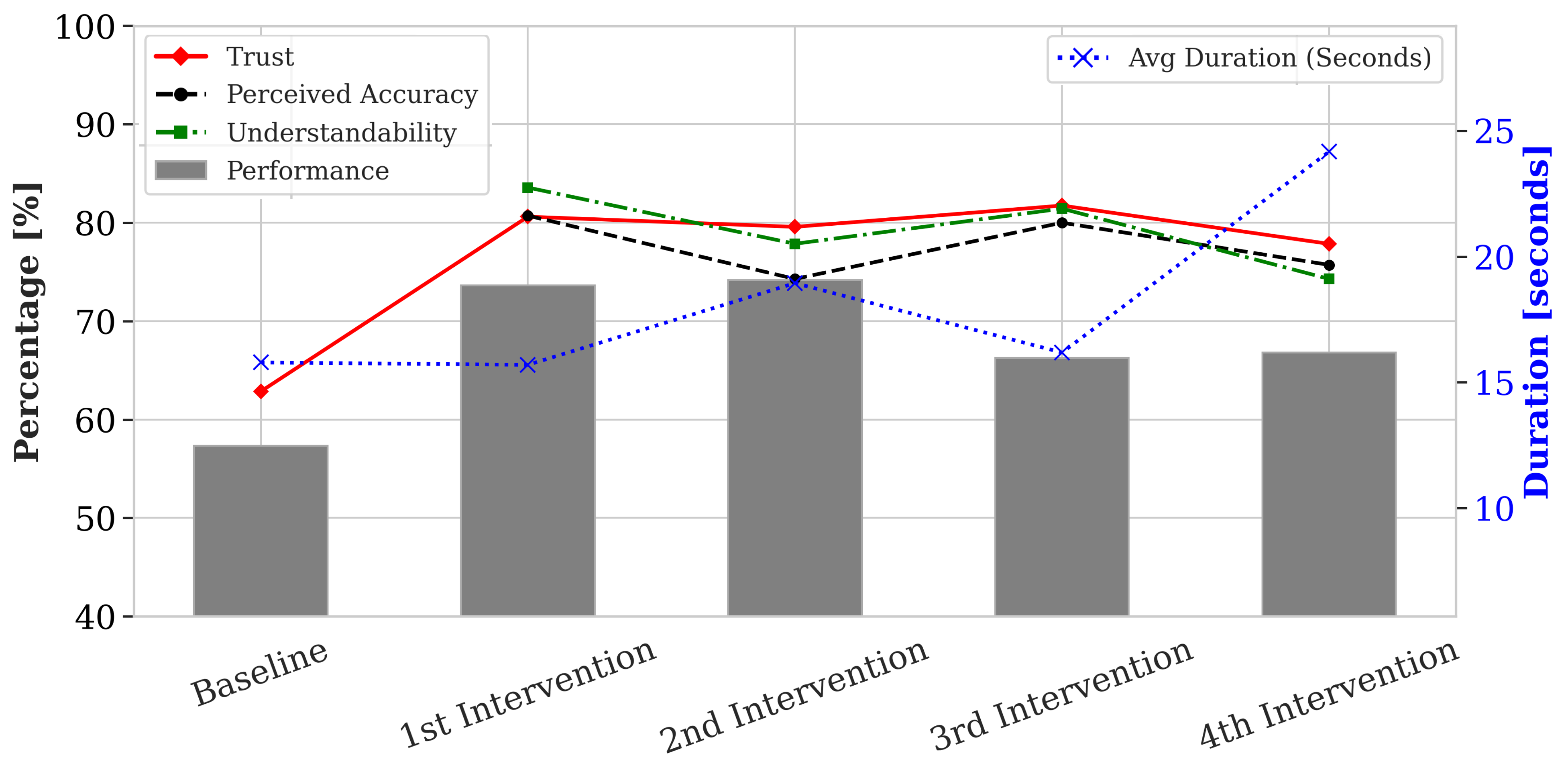}
    \caption{Summary of mean values of all key variables across interventions compared to the baseline with no AI}
    \label{fig:AVG}
\end{figure}

Analyzing all the results we presented (see a summary in \textit{\Cref{fig:AVG}}), highlights that using a CDSS regardless of its type provided notable improvements in performance and trust in clinical decision-making. On the other hand, increasing the level of explanations (all from 1st to 4th intervention) did not always lead to an increase in performance, trust or understandability. Furthermore, in some cases, particularly in the 4th intervention with highest level of explanations, we observed notable deterioration in all evaluation measures compared to the 1st intervention with least amount of explanations. Decision time in particular is the measure that was affected the most in the 4th intervention. Indeed, when all evaluation measures are considered holistically, 1st intervention with the simplest design would be the ideal choice among the four as seen in \textit{\Cref{fig:AVG}}).

Our findings present experimental evidence that support the value of CDSS in clinical decision-making. Particularly the improvement in diagnosis accuracy with CDSSs is remarkable. However, our results suggest that the designers should exercise caution when building explanations into CDSSs. The adverse effects of explanations as shown in our experiment may outweigh any potential benefits. 

\subsection{Limitations}
This study has several limitations that should be acknowledged. First, the controlled experimental setting does not fully reflect the complexity of real-world clinical environments. Participants were not interacting with real patients, nor were they working in a hospital setting, which means that crucial factors like emotional involvement, stress, and high-stakes decision-making were absent. Also, important issues such as workflow integration were outside the scope of this paper. The clinicians were interaction with a completely new system for a short duration during the experiment and implications of long-term use might be significantly different. Further research is necessary to the generalizability of our findings to actual clinical practice.

Additionally, our study relied on self-reported measures such as trust and workload which are subjective and may not fully capture participants' true behaviors or attitudes. The sample size and diversity of participants may also limit the broader applicability of the results, as our participant pool may not capture some effects such as cultural diversity found in clinical settings.

\section{Conclusion}
This paper presented an experimental study to test the impact of varying levels of AI explanations in CDSSs on the trust and diagnosis accuracy by 28 clinicians in a breast cancer detection problem. In this study we used a publicly available labelled breast cancer data set in an online user friendly interface to engage clinicians who hold various medical roles related to breast cancer diagnosis in a series of breast cancer diagnosis tasks supported by four different types of CDSSs with increasing level of explanations. In addition to the participant demographics, We evaluated self-reported measures of AI familiarity, trust, understandability, perceived accuracy as well as behavioral measures of diagnosis accuracy, agreement rate and diagnosis duration. Our results aim to provide guidance regrading how explanation mechanisms should be built into CDSSs with a cautionary note on the potential adverse effects of explanations by AI systems.

\bibliographystyle{ieeetr} 
\bibliography{referencesMod.bib}

\end{document}